# Solvent and Salt Effect on Lithium Ion Solvation and Contact Ion Pair Formation in Organic Carbonates: A Quantum Chemical Perspective

Veerapandian Ponnuchamy[†,‡,∥,*], Stefano Mossa[†,*] and Ioannis Skarmoutsos[†,*]

[†] Univ. Grenoble Alpes, CEA, CNRS, INAC-SYMMES, 38000 Grenoble, France

[‡] Univ. Grenoble Alpes, CEA, LITEN-DEHT, 38000 Grenoble, France

**Abstract**

Quantum chemical calculations have been employed to investigate the solvation of Lithium cations in ethylene carbonate / propylene carbonate and propylene carbonate / dimethyl carbonate mixed electrolytes. The impact of the presence of the counter anion on the solvation of $Li^+$ in pure propylene carbonate and dimethyl carbonate was also studied. The calculations revealed small free energy changes for the transitions between different preferred structures in mixed solvents. This implies that transitions between distinct local arrangements can take place in the mixtures. The addition of dimethyl carbonate causes a significant increase of the dipole moment of solvation clusters, indicating important molecular-scale modifications when dimethyl carbonate is used as co-solvent. The presence of an anion in the solvation shell of $Li^+$ modifies the intermolecular structure comprising four carbonate molecules in dilute solutions, allowing only two carbonate molecules to coordinate to $Li^+$. The bidentate complexation of $Li^+$ with the anion's electron donor atoms, however, maintains the local tetrahedral structure on interatomic length scales. The neutralization of the solvation shell of $Li^+$ due to contact ion-pair formation, and the consequent implications on the underlying mechanisms, provide a rational explanation for the ionic conductivity drop of electrolyte solutions at high salt concentrations.

[∥] Current address: InnoRenew CoE, Izola, Slovenia

* Emails: veera.pandi33@gmail.com / Phone: +33 7 78 10 42 94, stefano.mossa@cea.fr / Phone: +33 4 38 78 35 77 / Fax: +33 4 38 78 56 91, iskarmoutsos@hotmail.com / Phone: +33 4 38 78 35 77

## 1. Introduction

Rechargeable Lithium Ion Batteries (LIBs) are efficient and robust devices employed in various electronic applications, such as cellular phones and notebook computers, due to their high energy density and good reversibility[1-4]. Since the raise of LIBs in 1991, a tremendous amount of experimental and theoretical work has focused on investigating the chemistry involved in the utilized materials, in order to provide an efficient and environment-friendly system. A typical LIB consists of an anode (usually graphite) and a cathode (normally a transition metal oxide as, for example, $LiFeO_2$, $LiNiO_2$, $LiMn_2O_4$, or $LiCoO_2$)[5-10], separated by an electrolyte which acts as the ionic conducting medium and is typically composed by Lithium salts dissolved in non-aqueous organic carbonate solvents[1,11]. LIB performances are determined by the *combination* of the electrode materials, salts, and electrolyte solvents.

The choice of the electrolyte is capital, since it is involved in the Lithium ion transport, ultimately determines the conductivity, and controls the features (and quality) of the solid electrolyte interphase (SEI),[12-17] which forms on the electrode surface following the reductive decomposition of the electrolyte. This interphase strongly affects battery cyclability, lifetime, power and rate capabilities, and safety (flammability, for instance). Electrolytes employed in electrochemical devices must also be compatible with the electrodes, to ensure chemical stability, and exhibit a large electrochemical window between the lowest unoccupied and the highest occupied molecular orbitals (HOMO-LUMO), to provide thermochemical stability. Issues such as the temperature range for stability and the solvent efficiency in dissolving Lithium salts must also be optimized.

Phase behaviour modifications upon mixing are an additional important aspect. For instance, ethylene carbonate (EC) is solid at room temperature; however, mixed with other solvents such as linear carbonates or propylene carbonate (PC), it becomes liquid in a wide temperature range, with significant impact on energy storage features. Indeed, present most widely used electrolytes are mixtures of cyclic carbonates, such as EC and PC, with linear carbonates, as dimethyl carbonate (DMC). A wide range of LIB electrolyte solvents has been scrutinized following this route, searching for innovative mixtures exhibiting optimal values of key features, including high chemical stability, low melting point, high ionic conductivity, wide temperature range performance, and low viscosity.[18-23]. A substantial amount of evidences interestingly points to a crucial observation: The mechanisms underlying $Li^+$ ion transport are determined by a complex interplay among details of nanostructure and dynamics of the solvation sphere around the ion, collective solvent properties like the viscosity and dielectric permittivity, and ion pairs formation.

Although several experimental and theoretical works have focused on systems formed by *pure* EC, PC and DMC with the Li$^+$ ion[24-31], solvation structure and dynamics of the cation in these solvents is still a subject of debate. More in details, Density Functional Theory (DFT) studies have shown that a (four-coordinated) tetrahedral geometry Li$^+$(EC)$_4$, is the dominant species[32-34]. Ding et al.[35] performed DFT and Polarizable Continuum Model (PCM) studies to account for the coordination number around Li$^+$ ion, and suggested a transition between the most stable four-coordinated Li$^+$(EC)$_4$ and the three-coordinated trigonal Li$^+$(EC)$_3$ structures, due to a very low energy barrier separating the two geometries. On the other hand, Balbuena et al.[36] have suggested that Li$^+$(DMC)$_3$ is the most stable conformation in the Li$^+$-DMC system, while Li$^+$(PC)$_3$ is the leading component for the Li$^+$-PC case according to DFT studies of Balbuena et al.[32,37], Bhatt et al.[38] and Atetegeb et al.[39]. Gibbs Free Energies calculations in these latter studies also suggested that the tetra-coordinated Li$^+$(PC)$_4$ cluster is not favored in this case.

Balbuena et al.[36] performed DFT studies to investigate the solvation structure in various *mixtures* containing EC and linear carbonates (DMC, DEC). The reported results demonstrated that while the substitution of EC molecules by DMC or DEC in Li$^+$(EC)$_{2-3}$ clusters is not favourable, the addition of those linear molecules is still thermodynamically stable. Similarly, Bhatt et al.[40] investigated by DFT methods binary clusters consisting of EC and PC, with DMC and DEC as the second component, demonstrating that clusters with higher EC content are more stable. In the same work, it was also shown that the presence of linear carbonates breaks the symmetry of Li$^+$/EC clusters. Other important information includes the DFT studies of Klassen et al.[20] on Li$^+$/EC/PC clusters, which showed that the EC/Li$^+$ clusters have higher solvation energies than those formed by PC/Li$^+$, and EC selectively solvates Li$^+$. Similar DFT studies by Li et al.[33] also demonstrated that EC molecules can easily substitute PC in the first solvation shell of Li$^+$ in EC/PC mixtures.

The choice of the solvent is obviously not the only factor affecting Li$^+$ ion transport. Indeed, the presence of the *counter anions* included in the Li-salts (typically LiClO$_4$, LiBF$_4$ or LiPF$_6$) also influences the composition of the coordination shell and, therefore, modifies the dynamics of the dressed Li$^+$ ion. Raman spectroscopy studies[41] have shown that both EC and PC are bound to Li$^+$ in a 50:50 ratio in a 1M LiClO$_4$ system, thus indicating no preferential solvation of one species in the first solvation shell. Atomic Force Microscopy studies by Jeong et al.[42], on the other hand, revealed that in $LiClO_4$ solutions in mixed electrolytes, EC solvent molecules dominantly participate in the first solvation shell of $Li^+$ over linear carbonates. Molecular dynamics (MD) simulation studies of $LiBF_4$ in pure and mixed solvents[43], in addition, reported that $Li^+$ is preferentially solvated by the cyclic and more polar components in the mixtures, as the electrostatic interactions overcome possible steric hindrances. We observe, however, that in that work the estimated $Li^+$ coordination number was found to range from 5 to 6, depending on the salt concentration, a conclusion in contrast with the vast majority of existing theoretical and experimental studies.

In the case of $LiPF_6$ salt in mixed carbonate solvents, Morita et al.[44] performed Raman spectroscopic studies in EC/DMC mixtures, and suggested that EC coordination with $Li^+$ is preferred over the DMC one. Another $^{17}O$ NMR study[45], however, estimated a larger number of DMC molecules bound to $Li^+$ in the first solvation shell. In a combined DFT and MD study, Borodin et al.[46] found that the DFT optimized $Li^+(EC)_3DMC$ cluster is more stable than the $Li^+(EC)_4$, but classical and Born-Oppenheimer MD simulations[46,47] revealed a solvation shell with higher content of DMC than EC at high salt concentrations. Other classical MD simulation studies[48] suggested a preferential solvation of $Li^+$ by EC rather than DMC at low salt concentration, but very similar EC and DMC coordination numbers around $Li^+$ at high salt concentrations. In the case of $LiPF_6$ in EC/PC mixtures, the ab-initio MD studies of Ganesh et al.[49] concluded that EC solvates $Li^+$ more effectively than PC.

The above rapid (and partial) review clearly demonstrates both the complexity and the scope of the issues involved in a *general* description of the structure and preferential solvation properties of the $Li^+$ cations. It also points to two fundamental needs: first to verify systematically and on the same systems the compatibility of results coming from different numerical procedures employed at different length scales. Second, to clarify the relevance of the numerical results compared to experimental observations. In previous works[50,51] we moved on this line, with the goal of contributing general systematic information on these issues. We therefore focused on one side on a comparison of the outcomes of DFT calculations for small clusters and of classical MD simulations of larger ensembles of molecules[50]. On the other, we attempted at relating numerical data with novel experimental measurements able to provide a direct determination of the coordination structure of the $Li^+$ cations[51]. In the experimental-theoretical study of Ref. [51], in particular, based on femtosecond vibrational spectroscopy and DFT calculations we found that the EC coordination number around Li$^+$ cations strongly depends on salt concentration. For salt concentrations higher than 0.5 M, the coordination number of ethylene carbonate decreased to two, implying that both cation-anion interactions and ion-pair formation play a very important role on the structure and properties of the first solvation shell of Li$^+$. Interestingly, another recently published combined experimental and theoretical study[52] also pointed out that the coordination number of propylene carbonate and dimethyl carbonate in the first solvation shell of Li$^+$ can decrease from 5 to 2 when increasing the salt concentration. This further validates our previous studies and emphasizes the importance of ion-pair formation on the determination of the structure around Li$^+$ cations at high salt concentrations.

Here, we make a significant step forward in this direction, providing additional insight on the quantum mechanical features of the solvated clusters and the role played by the counter-ions. We first focus on the electronic structure, thermodynamic stability and vibrational properties of small clusters of the Li$^+$ ion in mixed EC/PC and PC/DMC solvents. Next, we extend our investigation with an analysis of the counter anion effects on the properties of the investigated clusters, with an emphasis on the (cation-anion) contact ion pair formation. We complete our work by relating our findings with previous theoretical and experimental studies.

## 2. Computational details

Following our previous works[50,51], quantum-chemical DFT calculations for several clusters of pure and binary electrolytes and counter anions around a lithium cation have been performed using the Amsterdam Density Functional (ADF) package[53]. In particular, the Perdew-Burke-Ernzerhof (PBE) GGA exchange correlation functional [54,55] was employed in our calculations, also using a TZP slater-type orbital (STO) basis set. Recent works[52] on systems similar to those investigated here also pointed out the increased accuracy of the PBE functional in predicting binding energies of solvated Li$^+$ complexes compared to possible alternatives like B3LYP or M05-2X. Our DFT calculations showed that the basis set superposition error (BSSE)[56] corrections are negligibly small, confirming a conclusion also reached in previous studies[34,57]. Vibrational frequency analysis were carried out for the optimized clusters, ensuring the absence of unstable (imaginary) modes and, therefore, confirming each structure as a minimum on the potential energy surface. Zero-Point Energy (ZPE) corrections have also been included. Thermodynamic quantities such as the entropy, enthalpy and Gibbs free energy were estimated at T=298.15 K, for all clusters.

## 3. Results and Discussion

### 3.1 Li$^+$ solvation in mixed electrolytes

In our previous studies[50] we found that the most thermodynamically stable complexes of solvent molecules around the Li$^+$ ion are tetra-coordinated. We also systematically studied the stability of the Li$^+$(EC)$_n$(DMC)$_m$, clusters, revealing that the thermodynamically most stable cluster is Li$^+$(EC)$_3$(DMC). On this basis, we conducted structural optimizations for the remaining clusters types containing two types of solvent molecules, Li$^+$(EC)$_n$(PC)$_m$ and Li$^+$(PC)$_n$(DMC)$_m$ with n+m=4. For the entire database of structures, we have also performed a vibrational frequency analysis and investigated charge transfer effects, comparing all data.

The Gibbs free energy changes have been estimated for the transitions of the form

$$Li^+(S_1)_k(S_2)_l + (4-l)\,S_1 + (4-k)\,S_2$$
$$\rightarrow Li^+(S_1)_n(S_2)_m + (4-n)\,S_1 + (4-m)\,S_2 \qquad (1)$$

where k+l=n+m=4. All possible combinations of the indexes were considered, in order to identify the most favourable cluster among those involved in the Li$^+$ coordination. The calculated binding energy (BE) values for the mixed composition EC/PC and PC/DMC clusters around Li$^+$, and the free energy values for the paths of the form $Li^+(S_1)_n(S_2)_m + (4-n)\,S_1 + (4-m)\,S_2$ are compared in Tables 1 and 2, respectively. The relative free energy values presented in Table 2 are rescaled by subtracting the minimum free energy value between the investigated fragments, and correspond to the free energy differences for the transition towards the most stable structure. The absolute values of the enthalpy and entropy of each cluster, as calculated with the ADF software, are provided in the Supporting Information (Table S1).

A few observations are in order. In general, the EC/PC/Li$^+$ clusters exhibit higher BE than those involving PC/DMC/Li$^+$. Among the EC/PC containing clusters, Li$^+$(EC)(PC)$_3$ exhibits the highest BE, but the total energy difference of these clusters are very small. It should also be noted that the BE values of Li$^+$/EC/PC mixtures are higher than those corresponding to clusters containing pure solvent molecules, such as Li$^+$(EC)$_4$ and Li$^+$(PC)$_4$ (BE is reported in Ref.[48]), confirming a trend already observed by Li et al.[33]. Among the PC/DMC/Li$^+$ clusters, Li$^+$(PC)$_3$(DMC) exhibits the highest BE value. The calculated Gibbs free energy values for EC/PC/Li$^+$ clusters also revealed that the binary cluster Li$^+$(EC)(PC)$_3$ is the most favourable, followed by Li$^+$(EC)$_2$(PC)$_2$ and, finally, by Li$^+$(EC)$_3$(PC). This finding is also in agreement with the experiments of Cresce et al.[58] complemented by ab initio calculations. In general, the trends in the binding energies are also similar with those obtained in Ref.[58], with the difference that the absolute value of the binding energy of Li$^+$(PC)$_4$ reported in our previous studies (-121.3 kcal/mol) is slightly lower than the one corresponding to Li$^+$(EC)(PC)$_3$. (This BE difference assumes, however, the relatively small value of 0.5 kcal/mol).

We also note that the Li$^+$(EC)(PC)$_3$ cluster is the most thermodynamically stable in the case of mixed EC/PC composition, but is less stable than Li$^+$(EC)$_4$. It should be also observed that the calculated free energy changes associated to the transitions between different types of structures are in general small. This indicates that, despite the existence of locally preferred structures, these small free energy changes could allow transitions between different solvation structures in the liquid state of EC/PC containing mixtures. In the case of PC/DMC/Li$^+$ clusters, Li$^+$(PC)$_2$(DMC)$_2$ and Li$^+$(PC)$_3$(DMC) exhibit similar free energy values indicating, again, that they can both exist in the bulk liquid phase. Interestingly, the least favourable cluster is the one with the highest concentration of DMC molecules, Li$^+$(PC)(DMC)$_3$. Note however that, as in the case of mixed EC/PC clusters, Li$^+$(PC)$_4$ is more stable than the mixed structures. These findings are also in agreement with recent experimental studies by Seo et al.[31].

The structural properties of Li$^+$(EC)$_n$(PC)$_m$ and Li$^+$(PC)$_n$(DMC)$_m$ clusters (with m+n=4) were also investigated. The optimized geometry bond parameters are reported in Table 3, together with the results obtained for the Li$^+$(EC)$_n$(DMC)$_m$ clusters, and the structures are shown in Figure 1(a)-(f). As we can see from Table 3, the variation in the relative proportion of EC or PC in clusters of the form Li$^+$(EC)$_n$(PC)$_m$ does not modify significantly the calculated bond distances, angles or dipole moment values, due to similar dipole moments and structures shared by EC and PC. In contrast, mixtures including DMC exhibit different bond lengths and angles, due to the substantially different structure and low dipole moment of DMC. Similar trends have also been reported in the MD studies of Lee et al.[59]. In these studies, molar volume calculations were performed, concluding that the PC/DMC mixtures are efficiently packed, due to the existence of attractive interactions, whereas repulsive interactions are more pronounced in EC/PC mixtures. These findings clearly underline that the presence of low-polar acyclic carbonates like DMC molecule in the first solvation shell breaks the higher order symmetry of EC/Li$^+$ and PC/Li$^+$ clusters. Interestingly, we can see from Table 3 that in the case of DMC molecules, the Li$^+$-O$_C$ distance is shorter than those corresponding to EC and PC. Moreover, the Li$^+$-O$_C$ distance for EC and PC is larger when DMC is present in the solvation shell.

In addition to the above structural properties, the atomic point charge distributions of the investigated clusters were obtained by performing a Multipole Derived Charge (MDC) analysis[60]. The MDC analysis uses the atomic multipoles (obtained from the fitted density) and reconstructs these multipoles exactly by distributing charges over all atoms. We report here our results, also including in the discussion the optimized Li$^+$(EC)$_n$(DMC)$_m$ clusters of Ref.[50]. The MDC point charges of the Li cation and the carbonyl oxygen O$_C$ do not exhibit significant variations with the composition of the first solvation sphere in the case of EC/PC mixtures. In DMC-containing clusters (EC/DMC and PC/DMC), however, stronger variations are evident. In general, in the cases where DMC is present in the first coordination shell, the corresponding carbonyl oxygen has a higher negative charge compared to the EC and PC cases, implying that the charge transfer effects between the Li cation and DMC molecules are stronger. Interestingly these charge transfer effects, combined with the structural symmetry breaking upon the addition of DMC, induce a significant increase of the total dipole moment of the cluster when DMC is present, as it can be seen in Table 3. This is in agreement with the findings of our previous classical MD simulation studies of Li$^+$ solvation in equimolar EC/DMC liquid mixtures[50]. All these results clearly emphasize the importance of the addition of DMC as a co-solvent in mixed cyclic-acyclic carbonate electrolytes for LIBs. We mention at this point that in our previous MD studies on dilute Li$^+$ solvation in a mixed EC-DMC solvent[50], where a flexible force field was used for DMC, we did not detect any conformational transition from the most stable cis-cis DMC$_{cc}$ conformer to other ones. As a consequence, in the present specific case of dilute Li$^+$ solvation clusters in mixed solvents containing DMC, where counter anions which could affect the DMC conformation are absent, only the most stable DMC$_{cc}$ was considered. Other issues related to the existence of conformers in pure DMC are discussed in the next section.

The effect of the mixed solvent composition on the vibrational properties was next addressed by calculating the vibrational spectra of the optimized clusters, shown in Figure 2(a)-(c). The vibrational frequencies associated to the C=O and O-C-O stretching modes are strongly affected by the coordination with the Li$^+$ ion in binary mixtures. In Figure 2(a), a sharp peak around 1800 cm$^{-1}$ and another feature around 1170 cm$^{-1}$ in EC/PC mixtures correspond to the C=O and O-C-O stretching vibrations, respectively. No significant changes of the C=O and O-C-O stretching frequencies at the indicated different ratios of EC/PC have been observed, indicating that the addition of PC to EC in mixed solvents does not strongly affect the interactions within the solvation shell around the lithium ion.

The addition of linear DMC solvent into the pure EC or PC electrolytes has, in contrast, a strong impact on vibrational frequencies. The corresponding IR spectra are shown in Figures 2(b) and (c), respectively. In particular, the C=O stretching frequency of EC or PC in EC/DMC and PC/DMC mixtures is red-shifted (of about 8 cm$^{-1}$) at higher DMC concentrations, while the O-C-O stretching frequency gradually increases. This finding is another clear manifestation of the changes taking place in the solvation shell of the Li cation upon the addition of DMC, another aspect motivating its importance as a co-solvent in mixed carbonate electrolytes. (Note that the calculated C=O stretching frequencies for PC and DMC molecules are in agreement with the experimental values reported by Seo et al[31] for coordinated PC and DMC values (Table S2 in the Supporting Information).)

### 3.2 Counter anion effects on Li$^+$ solvation

We now focus on the effects of the addition of a *counter anion* on the structural properties and the thermodynamic stability of clusters around the Li cation. (Note that counter anions are more likely to be present in the solvation shell of Li$^+$ at high salt concentrations. This is at variance with the results discussed above where we have considered solvation clusters which most likely mimic solvation at dilute lithium salt concentrations.) We have considered the counter anions ClO$_4^-$, BF$_4^-$ and PF$_6^-$, since they are the most frequently used lithium salts in LIBs. The gas phase DFT optimizations were performed for several pure PC and DMC solvent clusters around Li$^+$, in the form Li$^+$(S)$_{1-3}$ with S = PC and DMC. A similar analysis for EC is included in our Ref.[51]. We note also that our previous studies have revealed that the trends observed for gas-phase calculations do not change upon the addition of a dielectric continuum approximately mimicking the presence of a solvent.

The binding energies of Li$^+$ ion clusters containing solvent (S) molecules and the anion (A$^-$) were calculated as[53]

$$BE = E[Li^+(S)_{n=1-3}(A^-)] - E[Li^+] - E[A^-] - n\, E[S] \qquad (2)$$

Here, E[Li$^+$(S)$_{n=1-3}$ A$^-$] represents the total energy of the Li$^+$ ion cluster solvated by n solvent molecules with the counter anion nearby, E[Li$^+$] and E[A$^-$] are the energy of the isolated Li cation and the anion respectively, and E[S] is the energy of the isolated solvent molecule S.

The BE of the anion-containing PC and DMC clusters around Li$^+$ is presented in Table 4. From the obtained BE values (also calculated for the EC-containing clusters), we can conclude that the Li$^+$(DMC)(A$^-$) cluster exhibits higher binding energy than the corresponding EC and PC containing clusters. In particular, the values for the Li$^+$(DMC$_{cc}$)ClO$_4^-$, Li$^+$(EC)ClO$_4^-$ and Li$^+$(PC)ClO$_4^-$ clusters are -163.2, -163.1 and -162.6 kcal/mol, respectively. Similarly the corresponding values for A$^-$ = BF$_4^-$ clusters are -165.3, -163.8 and -165.3 kcal/mol, respectively, while those for A$^-$ = PF$_6^-$ are -158.7, -156.3 and 156.6 kcal/mol, respectively. Similar findings were already observed by Borodin et al.[46,47] in the case of PF$_6^-$ anion containing cluster, where the reported BE for Li$^+$(DMC)PF$_6^-$ was the highest one.

Following the gas-phase DFT energy optimizations for the Li$^+$(S)$_n$(A$^-$) clusters, based on the methodology employed in our previous studies[50,51], we calculated the free energy changes for fragments of the form Li$^+$(S)$_n$(A$^-$) + m S (here, again, S=PC and DMC, n + m = 3, and A$^-$ corresponds to the counter-anions ClO$_4^-$, BF$_4^-$ or PF$_6^-$). Subsequently we determined the free energy differences associated to the transitions

$$Li^+(S)_k A^- + l\, S \rightarrow Li^+(S)_n A^- + m\, S, \quad k+l = n+m = 3, \quad k,l,m,n = 0-3 \qquad (3)$$

In the cases where the above transitions exhibit negative free energy changes, this approach provides a clear indication that the formation of the Li$^+$(S)$_n$(A$^-$) cluster is favourable. The obtained relative free energy changes, shown in Table 5, reveal that the clusters including two solvent molecules and one counter-ion around the Li$^+$ ion are thermodynamically the most stable. Absolute values of the calculated thermodynamic properties of the clusters are provided in the Supporting Information (Table S3). The optimized most stable structures are shown in Figures 3, for PC ((a)-(c)) and DMC ((d)-(f)), respectively. Note that also *bulk* solvent effects have been studied, by applying a dielectric continuum in terms of the COSMO (COnductor-like Screening MOdel) model[61]. The calculated free energies values obtained by these calculations still show that Li$^+$(S)$_2$(A$^-$) is the most stable solvation cluster.

At this point we have to mention that in our recent combined experimental and DFT work[51], by using femtosecond vibrational spectroscopy it was estimated that at high salt concentration Li$^+$ is coordinated by about 2 ethylene carbonate solvent molecules in clusters containing the anion. Another very recent spectroscopic experimental study by Chapman et al[52] the solvation number decreases to values close to 2. In the case of the DMC solvent, the solvation number is even lower, due to the higher degree of association (see Figures 3b and 4b of Ref. [52]).

The conformation of DMC was also considered by performing calculations involving DMC (cis-trans) conformers. Our DFT calculations of the isolated cis-cis (DMCcc) and cis-trans (DMCct) conformers revealed that the DMCcc is relatively more stable than DMCct, with a free energy difference of about 3 kcal/mol. In addition, for consistency we extended our study for pure DMCct clusters in the presence of the anion Li+(DMCct)1-3(A-) in the first solvation shell. Thermodynamic properties of the DMCct containing clusters, along with the most stable optimized structures, are included in the Supporting Information (Tables S4, Figure S1). The results reveal the same trends for the solvation of DMC in the presence of a counter anion, where $Li^+(DMCct)_2 (A^-)$ is the most stable structure, as it can be seen in Table 5. Interestingly, the $Li^+ (DMCct)_{1-3}(A^-)$ clusters exhibit higher BE than those of the form $Li^+ (DMCcc)_{1-3}(A^-)$.

Also, the free energy changes associated to the transition from $Li^+(DMCcc)_2(A^-)$ to $Li^+(DMCcc)(DMCct)(A^-)$ and, finally, to $Li^+(DMCct)_2(A^-)$ ones (Table S5) are negative, implying that the relative stabilities of isolated DMC conformers are not crucial for the stabilization of solvation clusters. As a consequence, it is possible that different fractions of solvation structures containing different DMC conformers coexist in the condensed (bulk) phase. As pointed out in our previous study[50], however, only large scale Molecular Dynamics simulations of the liquid phase can provide more accurate information about these effects.

Also, from a visual inspection of the configurations shown in Figure 3 it is clear that a bidentate coordination is obtained in the cases of the most thermodynamically stable clusters, $Li^+(PC)_2(A^-)$ and $Li^+(DMC)_2(A^-)$. Similar conclusions were reached in our previous DFT study on $Li^+(EC)_n(Anion^-)$ clusters[51], which were also found to be in very good agreement with femtosecond vibrational spectroscopic measurements. These findings agree with previous studies[62-64] that also showed that at high salt concentration bi-dentate complexation of $Li^+$ with anions takes place.

Our data therefore show that, although the total molecular coordination number of $Li^+$ in the anion-containing first solvation reduces to three (including two solvent molecules and the anion), the Lithium cation is still closely interacting with four electron donor atoms. These features have two implications. First, the local molecular-scale tetrahedral structure comprising four carbonate solvent molecules around $Li^+$, widely accepted in dilute conditions, is strongly modified in concentrated salt solutions, leading to a first coordination shell of $Li^+$ comprising only two carbonate solvent molecules and one anion. Second, even in the presence of these strongly modified *molecular* environments, the $Li^+$ cation still keeps a tetrahedral coordination at the *interatomic* scale, interacting with the carbonyl oxygen atoms of the two carbonate solvents, and two electron donor atoms of the anions. Also, we note that, due to this specific structural arrangement inside the first solvation shell of $Li^+$, the charge of solvated $Li^+$ is neutralized by the anion and, as a consequence, we expect that its mobilty under an applied potential can be significantly reduced. The formation of such a cation/anion/solvent contact ion pair can therefore provide a rational explanation for the ionic conductivity drop of lithium salt /carbonate electrolyte solutions at high concentrations, as we will additionally discuss below.

We should mention at this point that long-range interactions could affect the solvation effects in condensed bulk phases, as we demonstrated in our combined classical MD-DFT studies of Ref.[50]. Note however, that the trends observed for the contact ion pairs remain the same even when a dielectric continuum is considered in the calculations. In any case, it should be pointed out that the investigation of long-range condensed phase effects on the determination of the relative fractions of contact ion-pairs, solvent-separated ions pairs and free ions at high salt concentrations is still an open issue. And these effects depend on several parameters including the polarity of the electrolyte, the nature of the anion or the concentration. To address these issues, combined information obtained from multiscale approaches are necessary. This goes well beyond the present work, whose target is to study solvation effects in contact ion pairs. In this framework, we provide insight on the mechanisms active at short intermolecular distances, and treat quantum chemically intra- and intermolecular interactions at these specific length scales.

The Gibbs free energy changes, ΔG, related to the solvation mechanism where two solvent molecules in the $Li^+$ first solvation shell are replaced by an anion have also been calculated, and are reported in Table 6. The highly negative ΔG values we have found indicate that this process, which can take place in electrolyte solutions with a high salt concentration resembling the conditions met in LIB devices, is highly spontaneous. From the magnitudes of the calculated ΔG values, it can also be observed that it is much easier to replace the low-dipolar $DMC_{cc}$ molecules (calculated dipole moment: 0.355 D) with an anion instead of PC or EC species. On the other hand, in the case of the $DMC_{ct}$ conformer, which in the gas phase has a much higher calculated dipole moment (3.50 D), the corresponding ΔG values are comparable (even slightly lower) to those pertaining to the polar EC and PC solvents. We note that previous studies[65] reported that only a few percent of cis-trans $DMC_{ct}$ conformers exist in the liquid state of the pure DMC solvent, and most of the molecules have the cis-cis $DMC_{cc}$ conformation, immediately justifying the low dielectric constant of the solvent. Our calculation also revealed that another DMC trans-trans conformer (with a calculated dipole moment of 5.03 D), $DMC_{tt}$, has a free energy which is 15.6 and 12.6 kcal/mol higher than that of $DMC_{cc}$ and $DMC_{ct}$, respectively. Due to these high free energy differences, this conformer was not considered in the present calculations. The finding that the ion pair formation is more pronounced in the low-polar $DMC_{cc}$ than in the highly-polar PC and EC solvents is also consistent with the trends in their dielectric constants[66,67], since ion pair formation is less pronounced in highly polar solvents. In addition, referring to the observed trends for the investigated anions, the substitution process is facilitated in the case of all solvents for $BF_4^-$, followed by $ClO_4^-$ and finally by $PF_6^-$. This finding hence additionally indicates that the contact ion pair formation is thermodynamically more favourable in the case of $LiBF_4$, followed by $LiClO_4$, with $LiPF_6$ being the least favourable case.

We conclude this discussion with an important observation. Previous works on electrical conductivity measurements[68,69] of $LiPF_6$, $LiClO_4$ and $LiBF_4$ solutions in organic carbonate electrolytes have reported general trends in the conductivity drop at high salt concentration similar to those described above in the contact ion pair formation. Indeed, according to the experimental measurements, the conductivity drop at high salt concentration is more pronounced for $LiBF_4$, followed by $LiClO_4$ and finally $LiPF_6$. In addition, the salt concentration corresponding to the maximum conductivity value is shown to increase following the trend $LiBF_4 < LiClO_4 < LiPF_6$ (Figure 4 in Ref.[68]). This finding is also consistent with our calculations since, in the cases where the contact ion pair formation is thermodynamically more pronounced, these effects should start manifesting at lower salt concentrations, since the probability to observe these ion pairs is in general higher.

Previous review works[1,70] have highlighted the fact that the behaviour of the conductivity in electrolyte solutions is a synergistic effect, controlled by both the viscosity of the solvent and the ion pair formation. It was also pointed out that the latter is probably the most important mechanism in the high salt concentration range. In contrast, in the opposite limit of low salt concentration, and until the point where the maximum conductivity value is observed, solvent viscosity and dielectric constant mainly control the conductivity. Our theoretical predictions, and their agreement with the experimental data, can therefore be considered as a further validation of the previously proposed mechanisms. From this point of view, they provide additional relevant insight that could be used as a springboard towards rational design of liquid electrolytes with optimal properties for battery applications.

## 4. Conclusions

Quantum chemical DFT calculations have been performed to clarify the solvation of lithium cations in mixed EC/PC and PC/DMC organic carbonate solvents, and investigate the counter anion effects on the solvation of these cations in pure PC and DMC solvents. These calculations have been integrated with those of our previous work[50,51] and the entire dataset has been rationalized in a single general discussion. More in details, a joint analysis of the changes in the free energy, associated to the transitions between different types of clusters formed by the lithium cation and the solvent molecules/anions, has been employed in order to identify the thermodynamically most stable clusters configurations forming the first solvation shell of $Li^+$. We summarize our main findings below.

In the case of the mixed EC/PC solvent, our analysis revealed that the binary cluster $Li^+(EC)(PC)_3$ is the most favourable one, followed by $Li^+(EC)_2(PC)_2$ and, finally, $Li^+(EC)_3(PC)$. However, the free energy changes between these clusters are very small, suggesting that any of them can possibly exist in EC/PC-containing mixtures. In the case of the mixed PC/DMC solvent, the free energy changes for the transition between the $Li^+(PC)_2(DMC)_2$ and $Li^+(PC)_3(DMC)$ clusters is very small, indicating that they can also both exist in the bulk liquid phase. In addition, the least favourable cluster is the one with the highest concentration of DMC molecules $Li^+(PC)(DMC)_3$.

The present study has also revealed that in the case of EC/PC mixtures, the charges of the Li cation and the carbonyl oxygen $O_C$ do not exhibit significant variations with the composition of the first solvation sphere. The only exception to this conclusion is the case of DMC containing clusters, which exhibit stronger modifications. We have found that in the cases where DMC is present in the first coordination shell, its corresponding carbonyl oxygen has a higher negative charge compared to the EC and PC cases, implying that charge transfers between the Li cation and DMC is stronger. These latter effects, combined with the structural symmetry breaking upon the addition of DMC, cause a significant increase of the dipole moment of the cluster. This is in agreement with the results of our previous classical MD simulation studies of $Li^+$ solvation in equimolar EC/DMC liquid mixtures[50], emphasizing the importance of the addition of DMC as a co-solvent in mixed cyclic-acyclic carbonate electrolytes for LIBs.

We also investigated the effect of the mixed solvent composition on vibrational properties, by calculating the vibrational spectra of the optimized clusters. The numerical data revealed that the vibrational frequencies associated to C=O stretching and O-C-O stretching of the carbonate molecules are strongly affected by the coordination with the $Li^+$ ion. Also, at variance with an increase of PC concentration in EC/PC mixed solvents which does not affect strongly the interactions within the solvation shell around $Li^+$, the addition of linear DMC solvent into the pure EC or PC electrolytes indeed strongly modifies their vibrational frequencies.

Finally, our study also highlighted an important point. The presence of an anion in the first coordination shell of $Li^+$ modifies the local tetrahedral structure, comprising four carbonate solvent molecules around $Li^+$ in dilute solutions, allowing two carbonate solvent molecules only to coordinate to $Li^+$ directly. However, a bidentate coordination is obtained in the cases of the most thermodynamically stable $Li^+(PC)_2(A^-)$ and $Li^+(DMC)_2(A^-)$ clusters, as also reported for $Li^+(EC)_2(A^-)$ clusters[51]. This finding indicates that although the total molecular coordination number of $Li^+$ in the anion-containing first solvation reduces to three, comprising two solvent molecules and one anion, the lithium cation is closely interacting with four electron donor atoms. As a consequence, due to this solvation structure the charge of solvated $Li^+$ is neutralized by the anion, and its mobilty under applied potential can be significantly reduced. This directly provides a rational explanation for the ionic conductivity drop of lithium salt /carbonate electrolyte solutions at high concentrations.

**Supporting Information**

Tables S1, S3, S4, S5. Indicated thermodynamic properties of the indicated solvation clusters. Table S2. Vibrational C=O stretching frequencies for PC and PMC. Figure S1. Optimized structures of the most stable contact ion-pair solvation structures containing cis-trans DMC conformers.

**Acknowledgments**

This work has been financially supported by the ANR-2011 PRGE002-04 ALIBABA and the DSM-Energie CEA Program.

**References**

(1) Xu, K. Nonaqueous Liquid Electrolytes for Lithium-Based Rechargeable Batteries. *Chem. Rev.* **2004**, *104*, 4303–4418.

(2) Armand, M.; Tarascon, J.-M. Building Better Batteries. *Nature* **2008**, *451*, 652–657.

(3) Palacín, M. R. Recent Advances in Rechargeable Battery Materials: A Chemist's Perspective. *Chem. Soc. Rev.* **2009**, *38*, 2565–2575.

(4) Goodenough, J. B.; Kim, Y. Challenges for Rechargeable Li Batteries. *Chem. Mater.* **2010**, *22*, 587–603.

(5) Armstrong, A. R.; Bruce, P. G. Synthesis of Layered $LiMnO_2$ as an Electrode for Rechargeable Lithium Batteries. *Nature* **1996**, *381*, 499–500.

(6) Birke, P.; Chu, W. F.; Weppner, W. Materials for Lithium Thin-Film Batteries for Application in Silicon Technology. *Solid State Ion.* **1996**, *93*, 1–15.

(7) Armstrong, A. R.; Tee, D. W.; La Mantia, F.; Novák, P.; Bruce, P. G. Synthesis of Tetrahedral $LiFeO_2$ and Its Behavior as a Cathode in Rechargeable Lithium Batteries. *J. Am. Chem. Soc.* **2008**, *130*, 3554–3559.

(8) Li, X.; Cheng, F.; Guo, B.; Chen, J. Template-Synthesized $LiCoO_2$, $LiMn_2O_4$, and $LiNi_{0.8}Co_{0.2}O_2$ Nanotubes as the Cathode Materials of Lithium Ion Batteries. *J. Phys. Chem. B* **2005**, *109*, 14017–14024.

(9) Cho, J.; Kim, Y. J.; Park, B. Novel $LiCoO_2$ Cathode Material with $Al_2O_3$ Coating for a Li Ion Cell. *Chem. Mater.* **2000**, *12*, 3788–3791.

(10) Bhatt, M. D.; O'Dwyer, C. Recent Progress in Theoretical and Computational Investigations of Li-ion Battery Materials and Electrolytes. *Phys. Chem. Chem. Phys.* **2015**, *17*, 4799–4844.

(11) Xu, K. Electrolytes and Interphases in Li-Ion Batteries and Beyond. *Chem. Rev.* **2014**, *114*, 11503–11618.

(12) Peled, E. The Electrochemical Behavior of Alkali and Alkaline Earth Metals in Nonaqueous Battery Systems – The Solid Electrolyte Interphase Model. *J. Electrochem. Soc.* **1979**, *126*, 2047–2051.

(13) Peled, E.; Golodnitsky, D.; Ardel, G. Advanced Model for Solid Electrolyte Interphase Electrodes in Liquid and Polymer Electrolytes. *J. Electrochem. Soc.* **1997**, *144*, L208–L210.

(14) Arora, P.; White, R. E.; Doyle, M. Capacity Fade Mechanisms and Side Reactions in Lithium Ion Batteries. *J. Electrochem. Soc.* **1998**, *145*, 3647–3667.

(15) Verma, P.; Maire, P.; Novák, P. A Review of the Features and Analyses of the Solid Electrolyte Interphase in Li-Ion Batteries. *Electrochimica Acta* **2010**, *55*, 6332–6341.

(16) Muralidharan, A.; Chaudhari, M.I.; Pratt, L.R.; Rempe, S.B. Molecular Dynamics of Lithium Ion Transport in a Model Solid Electrolyte Interphase. *Sci. Rep.* **2018**, *8*, 10736.

(17) Leung, K.; Budzien, J.L. Ab initio Molecular Dynamics Simulations of the Initial Stages of Solid–Electrolyte Interphase Formation on Lithium ion Battery Graphitic Anodes. *Phys. Chem. Chem. Phys.* **2010**, *12*, 6583-6586.

(18) Tobishima, S.; Yamaji, A. Electrolytic Characteristics of Mixed Solvent Electrolytes for Lithium Secondary Batteries. *Electrochimica Acta* **1983**, *28*, 1067–1072.

(19) Tarascon, J. M.; Guyomard, D. New Electrolyte Compositions Stable over the 0 to 5 V Voltage Range and Compatible with the $Li_{1+x}Mn_2O_4$/Carbon Li-Ion Cells. *Solid State Ion.* **1994**, *69*, 293–305.

(20) Klassen, B.; Aroca, R.; Nazri, M.; Nazri, G. A. Raman Spectra and Transport Properties of Lithium Perchlorate in Ethylene Carbonate Based Binary Solvent Systems for Lithium Batteries. *J. Phys. Chem. B* **1998**, *102*, 4795–4801.

(21) Matsuda, Y.; Fukushima, T.; Hashimoto, H.; Arakawa, R. Solvation of Lithium Ions in Mixed Organic Electrolyte Solutions by Electrospray Ionization Mass Spectroscopy. *J. Electrochem. Soc.* **2002**, *149*, A1045–A1048.

(22) Eshetu, G. G.; Bertrand, J.-P.; Lecocq, A.; Grugeon, S.; Laruelle, S.; Armand, M.; Marlair, G. Fire Behavior of Carbonates-Based Electrolytes Used in Li-Ion Rechargeable Batteries with a Focus on the Role of the $LiPF_6$ and LiFSI Salts. *J. Power Sources* **2014**, *269*, 804–811.

(23) Yuan, K.; Bian, H.; Shen, Y.; Jiang, B.; Li, J.; Zhang, Y.; Chen, H.; Zheng, J. Coordination Number of $Li^+$ in Nonaqueous Electrolyte Solutions Determined by Molecular Rotational Measurements. *J. Phys. Chem. B* **2014**, *118*, 3689–3695.

(24) Tsunekawa, H.; Narumi, A.; Sano, M.; Hiwara, A.; Fujita, M.; Yokoyama, H. Solvation and Ion Association Studies of $LiBF_4$−Propylenecarbonate and $LiBF_4$−Propylenecarbonate−Trimethyl Phosphate Solutions. *J. Phys. Chem. B* **2003**, *107*, 10962–10966.

(25) Castriota, M.; Cazzanelli, E.; Nicotera, I.; Coppola, L.; Oliviero, C.; Ranieri, G. A. Temperature Dependence of Lithium Ion Solvation in Ethylene Carbonate–$LiClO_4$ Solutions. *J. Chem. Phys.* **2003**, *118*, 5537–5541.

(26) Kameda, Y.; Umebayashi, Y.; Takeuchi, M.; Wahab, M. A.; Fukuda, S.; Ishiguro, S.; Sasaki, M.; Amo, Y.; Usuki, T. Solvation Structure of $Li^+$ in Concentrated $LiPF_6$−Propylene Carbonate Solutions. *J. Phys. Chem. B* **2007**, *111*, 6104–6109.

(27) Xu, K.; Lam, Y.; Zhang, S. S.; Jow, T. R.; Curtis, T. B. Solvation Sheath of $Li^+$ in Nonaqueous Electrolytes and Its Implication of Graphite/Electrolyte Interface Chemistry. *J. Phys. Chem. C* **2007**, *111*, 7411–7421.

(28) W. Smith, J.; K. Lam, R.; T. Sheardy, A.; Shih, O.; M. Rizzuto, A.; Borodin, O.; J. Harris, S.; Prendergast, D.; J. Saykally, R. X-Ray Absorption Spectroscopy of $LiBF_4$ in Propylene Carbonate: A Model Lithium Ion Battery Electrolyte. *Phys. Chem. Chem. Phys.* **2014**, *16*, 23568–23575.

(29) Seo, D. M.; Reininger, S.; Kutcher, M.; Redmond, K.; Euler, W. B.; Lucht, B. L. Role of Mixed Solvation and Ion Pairing in the Solution Structure of Lithium Ion Battery Electrolytes. *J. Phys. Chem. C* **2015**, *119*, 14038–14046.

(30) Chaudhari, M.I.; Nair, J.R.; Pratt, L.R.; Soto, F.A.; Balbuena, P.B.; Rempe, S.B. Scaling Atomic Partial Charges of Carbonate Solvents for Lithium Ion Solvation and Diffusion. *J. Chem. Theory Comput.* **2016**, *12*, 5709–5718.

(31) Chaudhari, M.I.; Muralidharan, A.; Pratt, L.R.; Rempe, S.B. Assessment of Simple Models for Molecular Simulation of Ethylene Carbonate and Propylene Carbonate as Solvents for Electrolyte Solutions. *Top. Curr. Chem. (Z)* **2018**, *376*:7.

(32) Wang, Y.; Nakamura, S.; Ue, M.; Balbuena, P. B. Theoretical Studies to Understand Surface Chemistry on Carbon Anodes for Lithium-Ion Batteries: Reduction Mechanisms of Ethylene Carbonate. *J. Am. Chem. Soc.* **2001**, *123*, 11708–11718.

(33) Li, T.; Balbuena, P. B. Theoretical Studies of Lithium Perchlorate in Ethylene Carbonate, Propylene Carbonate, and Their Mixtures. *J. Electrochem. Soc.* **1999**, *146*, 3613–3622.

(34) Bhatt, M. D.; Cho, M.; Cho, K. Interaction of Li$^+$ Ions with Ethylene Carbonate (EC): Density Functional Theory Calculations. *Appl. Surf. Sci.* **2010**, *257*, 1463–1468.

(35) Ding, W.; Lei, X.; Ouyang, C. Coordination of Lithium Ion with Ethylene Carbonate Electrolyte Solvent: A Computational Study. *Int. J. Quantum Chem.* **2016**, *116*, 97–102.

(36) Wang, Y.; Balbuena, P. B. Theoretical Studies on Cosolvation of Li Ion and Solvent Reductive Decomposition in Binary Mixtures of Aliphatic Carbonates. *Int. J. Quantum Chem.* **2005**, *102*, 724–733.

(37) Wang, Y.; Balbuena, P. B. Theoretical Insights into the Reductive Decompositions of Propylene Carbonate and Vinylene Carbonate: Density Functional Theory Studies. *J. Phys. Chem. B* **2002**, *106*, 4486–4495.

(38) Bhatt, M. D.; Cho, M.; Cho, K. Conduction of Li$^+$ Cations in Ethylene Carbonate (EC) and Propylene Carbonate (PC): Comparative Studies Using Density Functional Theory. *J. Solid State Electrochem.* **2012**, *16*, 435–441.

(39) Haregewoin, A. M.; Leggesse, E. G.; Jiang, J.-C.; Wang, F.-M.; Hwang, B.-J.; Lin, S. D. A Combined Experimental and Theoretical Study of Surface Film Formation: Effect of Oxygen on the Reduction Mechanism of Propylene Carbonate. *J. Power Sources* **2013**, *244*, 318–327.

(40) Bhatt, M. D.; O'Dwyer, C. Density Functional Theory Calculations for Ethylene Carbonate-Based Binary Electrolyte Mixtures in Lithium Ion Batteries. *Curr. Appl. Phys.* **2014**, *14*, 349–354.

(41) Hyodo, S.-A.; Okabayashi, K. Raman Intensity Study of Local Structure in Non-Aqueous Electrolyte Solutions—II. Cation—solvent Interaction in Mixed Solvent Systems and Selective Solvation. *Electrochimica Acta* **1989**, *34*, 1557–1561.

(42) Jeong, S.-K.; Inaba, M.; Mogi, R.; Iriyama, Y.; Abe, T.; Ogumi, Z. Surface Film Formation on a Graphite Negative Electrode in Lithium-Ion Batteries: Atomic Force Microscopy Study on the Effects of Film-Forming Additives in Propylene Carbonate Solutions. *Langmuir* **2001**, *17*, 8281–8286.

(43) Postupna, O. O.; Kolesnik, Y. V.; Kalugin, O. N.; Prezhdo, O. V. Microscopic Structure and Dynamics of LiBF$_4$ Solutions in Cyclic and Linear Carbonates. *J. Phys. Chem. B* **2011**, *115*, 14563–14571.

(44) Morita, M.; Asai, Y.; Yoshimoto, N.; Ishikawa, M. A Raman Spectroscopic Study of Organic Electrolyte Solutions Based on Binary Solvent Systems of Ethylene Carbonate with Low Viscosity Solvents Which Dissolve Different Lithium Salts. *J. Chem. Soc. Faraday Trans.* **1998**, *94*, 3451–3456.

(45) Bogle, X.; Vazquez, R.; Greenbaum, S.; Cresce, A. von W.; Xu, K. Understanding Li$^+$–Solvent Interaction in Nonaqueous Carbonate Electrolytes with $^{17}$O NMR. *J. Phys. Chem. Lett.* **2013**, *4*, 1664–1668.

(46) Borodin, O.; Smith, G. D. Quantum Chemistry and Molecular Dynamics Simulation Study of Dimethyl Carbonate: Ethylene Carbonate Electrolytes Doped with LiPF$_6$. *J. Phys. Chem. B* **2009**, *113*, 1763–1776.

(47) Borodin, O.; Olguin, M.; Ganesh, P.; Kent, P. R.; Allen, J. L.; Henderson, W. A. Competitive Lithium Solvation of Linear and Cyclic Carbonates from Quantum Chemistry. *Phys. Chem. Chem. Phys* **2016**, *18*, 164-175.

(48) Tenney, C. M.; Cygan, R. T. Analysis of Molecular Clusters in Simulations of Lithium-Ion Battery Electrolytes. *J. Phys. Chem. C* **2013**, *117*, 24673–24684.

(49) Ganesh, P.; Jiang, D.; Kent, P. R. C. Accurate Static and Dynamic Properties of Liquid Electrolytes for Li-Ion Batteries from Ab Initio Molecular Dynamics. *J. Phys. Chem. B* **2011**, *115*, 3085–3090.

(50) Skarmoutsos, I.; Ponnuchamy, V.; Vetere, V.; Mossa, S. Li$^+$ Solvation in Pure, Binary, and Ternary Mixtures of Organic Carbonate Electrolytes. *J. Phys. Chem. C* **2015**, *119*, 4502–4515.

(51) Jiang, B.; Ponnuchamy, V.; Shen, Y.; Yang, X.; Yuan, K.; Vetere, V.; Mossa, S.; Skarmoutsos, I.; Zhang, Y.; Zheng, J. The Anion Effect on Li$^+$ Ion Coordination Structure in Ethylene Carbonate Solutions. *J. Phys. Chem. Lett.* **2016**, *7*, 3554–3559.

(52) Chapman, N.; Borodin, O.; Yoon, T.; Nguyen, C. C.; Lucht, B. L. Spectroscopic and Density Functional Theory Characterization of Common Lithium Salt Solvates in Carbonate Electrolytes for Lithium Batteries. *J. Phys. Chem. C* **2017**, *121*, 2135-2148.

(53) te Velde, G.; Bickelhaupt, F. M.; Baerends, E. J.; Fonseca Guerra, C.; van Gisbergen, S. J. A.; Snijders, J. G.; Ziegler, T. Chemistry with ADF. *J. Comput. Chem.* **2001**, *22*, 931–967.

(54) Perdew, J. P.; Burke, K.; Ernzerhof, M. Generalized Gradient Approximation Made Simple. *Phys. Rev. Lett.* **1996**, *77*, 3865–3868.

(55) Tasaki, K.; Goldberg, A.; Winter, M. On the Difference in Cycling Behaviors of Lithium-Ion Battery Cell between the Ethylene Carbonate- and Propylene Carbonate-Based Electrolytes. *Electrochimica Acta* **2011**, *56*, 10424–10435.

(56) Boys, S. F.; Bernardi, F. The Calculation of Small Molecular Interactions by the Differences of Separate Total Energies. Some Procedures with Reduced Errors. *Mol. Phys.* **1970**, *19*, 553–566.

(57) Bhatt, M. D.; Cho, M.; Cho, K. Density Functional Theory Calculations and Ab Initio Molecular Dynamics Simulations for Diffusion of Li$^+$ within Liquid Ethylene Carbonate. *Model. Simul. Mater. Sci. Eng.* **2012**, *20*, 065004.

(58) von Wald Cresce, A.; Borodin, O.; Xu, K. Correlating Li$^+$ Solvation Sheath Structure with Interphasial Chemistry on Graphite. *J. Phys. Chem. C* **2012**, *116*, 26111-26117.

(59) Lee, S.; Park, S. S. Thermodynamic and Dynamic Properties in Binary Mixtures of Propylene Carbonate with Dimethyl Carbonate and Ethylene Carbonate. *J. Mol. Liq.* **2012**, *175*, 97–102.

(60) Swart, M.; van Duijnen, P.; Snijders, J. A Charge Analysis Derived from an Atomic Multipole Expansion. *J. Comput. Chem.* **2001**, *22*, 79−88.

(61) Klamt, A.; Schüürman, G. COSMO: A New Approach to Dielectric Screening in Solvents with Explicit Expressions for the Screening Energy and its Gradient. *J. Chem. Soc., Perkin Trans. 2* **1993**, *5*, 799-805.

(62) Flores, E.; Aval, G.; Jeschke, S.; Johansson, P. Solvation Structure in Dilute to Highly Concentrated Electrolytes for Lithium-ion and Sodium-ion Batteries. *Electrochimica Acta* **2017**, *233*, 134-141.

(63) Maeda, S.; Kameda, Y.; Amo, Y.; Usuki, T.; Ikeda, K.; Otomo, T.; Yanagisawa, M.; Seki, S.; Arai, N.; Watanabe, H.; Umebayashi, Y. Local Structure of Li$^+$ in Concentrated Ethylene Carbonate Solutions Studied by Low-frequency Raman Scattering and Neutron Diffraction with $^6$Li/$^7$Li Isotopic Substitution Methods. *J. Phys. Chem. B* **2017**, *121*, 10979–10987.

(64) Klassen, B.; Aroca, R. Lithium Perchlorate: Ab Initio Study of the Structural and Spectral Changes Associated with Ion Pairing. *J. Phys. Chem.* **1996**, *100*, 9334–9338.

(65) Reddy, S.K.; Balasubramanian, S. Liquid Dimethyl Carbonate: A Quantum Chemical and Molecular Dynamics Study. *J. Phys. Chem. B* **2012**, *116*, 14892–14902.

(66) Payne, R.; Theodorou, I.E. Dielectric Properties and Relaxation in Ethylene Carbonate and Propylene Carbonate. *J. Phys. Chem.* **1972**, *76*, 2892-2900.

(67) Naejus, R.; Lemordant, D.; Coudert, R.; Willman P. Excess Thermodynamic of Binary Mixtures Containing Linear or Cyclic Carbonates as Solvents at the Temperatures 298.15 K and 313.15 K. *J. Chem. Thermodyn.* **1997**, *29*, 1503-1515.

(68) Moumouzias, G.; Ritzoulis, G.; Siapkas, D.; Terzidis, D. Comparative Study of LiBF$_4$, LiAsF$_6$, LiPF$_6$ and LiClO$_4$ as Electrolytes in Propylene Carbonate-Diethyl Carbonate Solutions for Li/LiMn$_2$O$_4$ Cells. *J. Power Sources* **2003**, *122*, 57-66.

(69) Takeuchi, M.; Kameda, Y.; Umebayashi, Y.; Ogawa, S.; Sonoda, T.; Ishiguro, S-I.; Fujita, M.; Sano, M. Ion-ion Interactions of LiPF$_6$ and LiBF$_4$ in Propylene Carbonate Solutions. *J. Mol. Liq.* **2009**, *148*, 99-108.

(70) Marcus, Y.; Hefter, G. Ion Pairing. *Chem. Rev.* **2006**, *106*, 4585-4621.

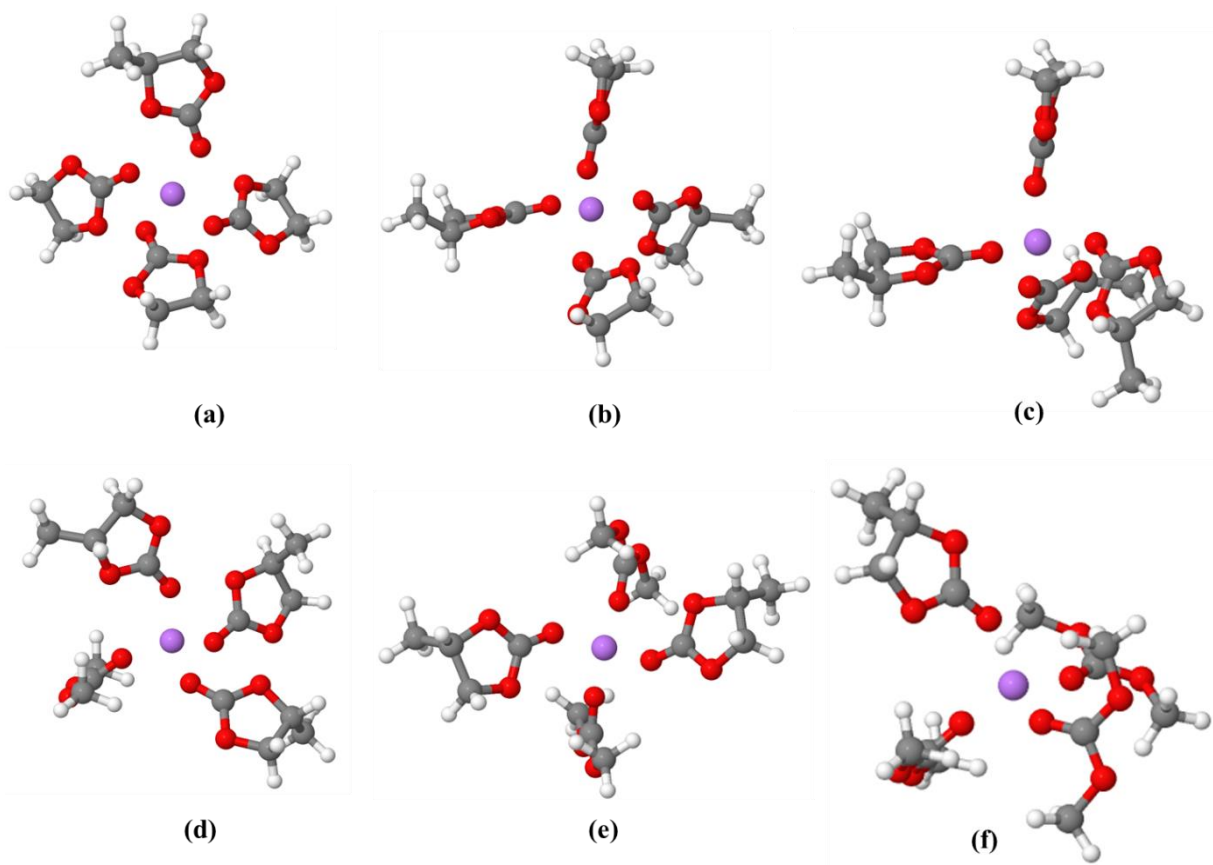

**Figure 1**. Optimized geometries of (a) Li$^+$(EC)$_3$(PC), (b) Li$^+$(EC)$_2$(PC)$_2$, (c) Li$^+$(EC)(PC)$_3$, (d) Li$^+$(PC)$_3$(DMC), (e) Li$^+$(PC)$_2$(DMC)$_2$, and (f) Li$^+$(PC)(DMC)$_3$ clusters.

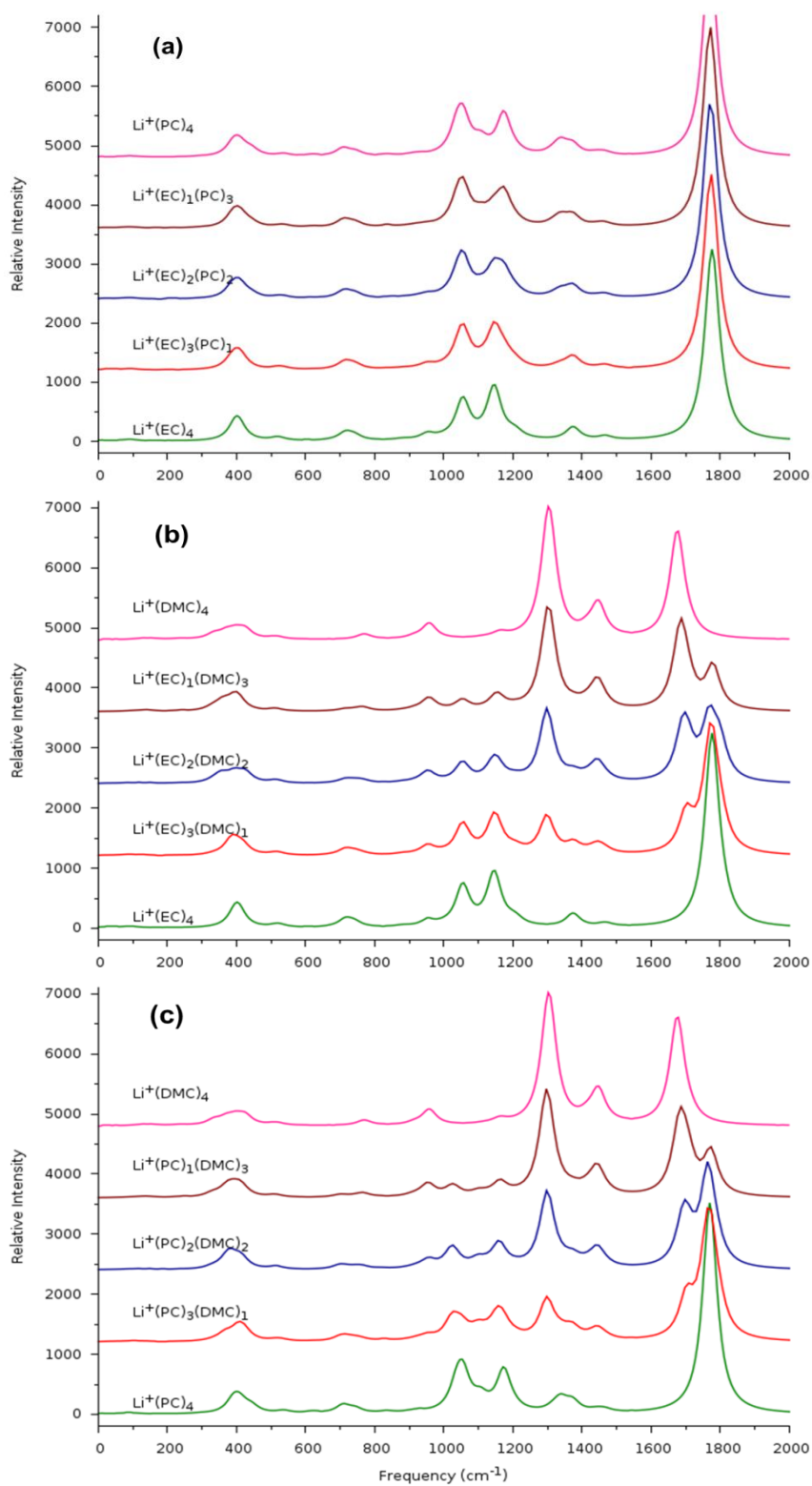

**Figure 2**. Comparison of the IR spectra for the investigated clusters of the form (a) $Li^+(EC)_n(PC)_m$, (b) $Li^+(EC)_n(DMC)_m$ and (c) $Li^+(PC)_n(DMC)_m$, with n+m =4. The relative compositions are indicated in all cases.

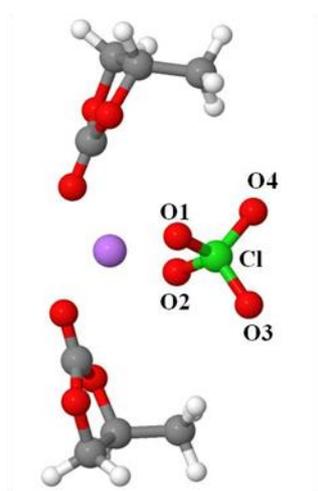 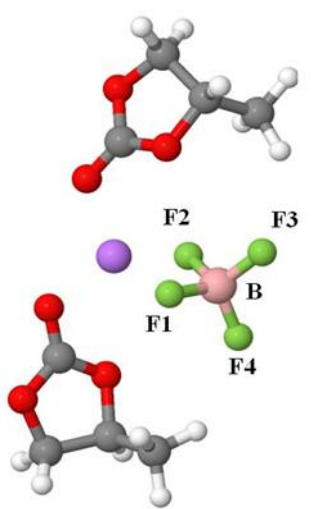 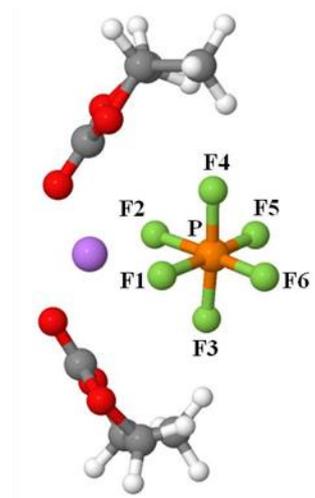
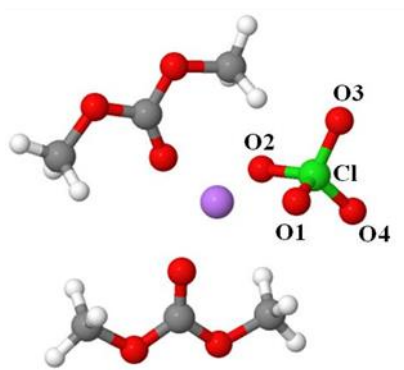 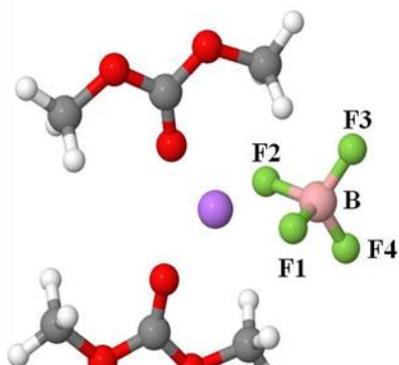 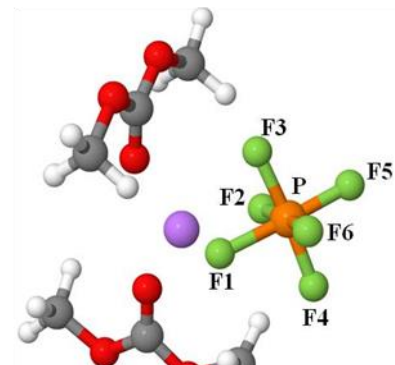

**Figure 3**. Most stable optimized geometries of (a) Li$^+$(PC)$_2$(ClO$_4^-$), (b) Li$^+$(PC)$_2$(BF$_4^-$), (c) Li$^+$(PC)$_2$(PF$_6^-$), (d) Li$^+$(DMC)$_2$(ClO$_4^-$), (e) Li$^+$(DMC)$_2$(BF$_4^-$), and (f) Li$^+$(DMC)$_2$(PF$_6^-$).

**Table 1.** Binding Energy (BE) of $Li^+(EC)_n(PC)_m$ and $Li^+(PC)_n(DMC)_m$ (n + m = 4) clusters

| Clusters | BE (kcal/mol) |
|---|---|
| $Li^+(EC)_3(PC)$ | -120.5 |
| $Li^+(EC)_2(PC)_2$ | -121.2 |
| $Li^+(EC)(PC)_3$ | -121.8 |
| $Li^+(PC)_3(DMC_{cc})$ | -118.4 |
| $Li^+(PC)_2(DMC_{cc})_2$ | -114.5 |
| $Li^+(PC)(DMC_{cc})_3$ | -110.2 |

**Table 2.** Relative Gibbs Free Energies ΔG (in kcal/mol) of Li$^+$(EC)$_n$(PC)$_m$ and Li$^+$(PC)$_n$(DMC$_{cc}$)$_m$ (with n + m = 4) containing fragments.

| Fragments | ΔG=G-G$^{min}$ * (kcal/mol) |
|---|---|
| Li$^+$(EC)$_3$(PC) + EC + 3 PC | 2.0 |
| Li$^+$(EC)$_2$(PC)$_2$ + 2 EC + 2 PC | 1.3 |
| Li$^+$(EC)(PC)$_3$ + 3 EC + PC | 0.8 |
| Li$^+$(EC)$_4$ + 4 PC | 0.0 |
| Li$^+$(PC)$_4$ + 4 EC | 1.6 |
| | |
| Li$^+$(PC)$_3$(DMC$_{cc}$) + PC + 3 DMC$_{cc}$ | 1.6 |
| Li$^+$(PC)$_2$(DMC$_{cc}$)$_2$ + 2 PC + 2 DMC$_{cc}$ | 1.5 |
| Li$^+$(PC)(DMC$_{cc}$)$_3$ + 3 PC + DMC$_{cc}$ | 8.1 |
| Li$^+$(PC)$_4$ + 4 DMC$_{cc}$ | 0.0 |
| Li$^+$(DMC$_{cc}$)$_4$ + 4 PC | 13.2 |

\* Gmin corresponds to the minimum free energy value for each of the 2 separate set data presented in the 2 different columns

**Table 3.** Structural Parameters of Optimized Binary Clusters (Bond Lengths in Angstrom, Bond Angles in degree), together with Dipole (Debye) and Multipole-Derived Charges (e)

| Mixtures | $r_{Li+-O}$ | | $r_{O=C}$ | | Angle $_{Li+-O=C}$ | | Dipole | Charges ($Li^{\delta+}$, $O_C$) | | |
|---|---|---|---|---|---|---|---|---|---|---|
| | $S_1$ | $S_2$ | $S_1$ | $S_2$ | $S_1$ | $S_2$ | | $Li^{\delta+}$ | $O_C(S_1)$ | $O_C(S_2)$ |
| $Li^+$: $EC(S_1)$ : $PC(S_2)$ | | | | | | | | | | |
| 1:3:1 | 1.953 | 1.948 | 1.216 | 1.218 | 135.9 | 134.2 | 0.44 | 0.842 | -0.474 | -0.455 |
| 1:2:2 | 1.955 | 1.949 | 1.216 | 1.217 | 136.0 | 134.0 | 0.50 | 0.850 | -0.481 | -0.434 |
| 1:1:3 | 1.957 | 1.951 | 1.215 | 1.217 | 136.5 | 136.2 | 0.40 | 0.856 | -0.448 | -0.434 |
| $Li^+$: $EC(S_1)$: $DMC_{cc}(S_2)$ | | | | | | | | | | |
| 1:3:1 | 1.943 | 1.921 | 1.216 | 1.231 | 138.9 | 160.8 | 7.26 | 0.810 | -0.397 | -0.462 |
| 1:2:2 | 1.972 | 1.910 | 1.214 | 1.231 | 136.6 | 156.4 | 8.30 | 0.743 | -0.318 | -0.494 |
| 1:1:3 | 1.943 | 1.921 | 1.216 | 1.231 | 138.9 | 160.8 | 6.51 | 0.688 | -0.221 | -0.481 |
| $Li^+$: $PC(S_1)$: $DMC_{cc}(S_2)$ | | | | | | | | | | |
| 1:3:1 | 1.951 | 1.919 | 1.215 | 1.231 | 147.3 | 152.6 | 7.82 | 0.809 | -0.345 | -0.416 |
| 1:2:2 | 1.970 | 1.912 | 1.217 | 1.231 | 140.4 | 164.6 | 5.55 | 0.778 | -0.271 | -0.479 |
| 1:1:3 | 1.954 | 1.950 | 1.219 | 1.233 | 148.4 | 157.1 | 6.29 | 0.725 | -0.341 | -0.438 |

**Table 4.** Binding Energies (BE) of Li$^+$(PC)$_{1-3}$(A$^-$) and Li$^+$(DMC)$_{1-3}$(A$^-$) Clusters

| Fragments | BE (kcal/mol) |
|---|---|
| Li$^+$(PC)ClO$_4^-$ | -162.6 |
| Li$^+$(PC)$_2$ClO$_4^-$ | -176.0 |
| Li$^+$(PC)$_3$ClO$_4^-$ | -185.3 |
| Li$^+$(PC)BF$_4^-$ | -165.3 |
| Li$^+$(PC)$_2$BF$_4^-$ | -179.4 |
| Li$^+$(PC)$_3$BF$_4^-$ | -189.3 |
| Li$^+$(PC)PF$_6^-$ | -156.6 |
| Li$^+$(PC)$_2$PF$_6^-$ | -173.5 |
| Li$^+$(PC)$_3$PF$_6^-$ | -181.5 |
| | |
| Li$^+$(DMC$_{cc}$)ClO$_4^-$ | -163.2 |
| Li$^+$(DMC$_{cc}$)$_2$ClO$_4^-$ | -176.3 |
| Li$^+$(DMC$_{cc}$)$_3$ClO$_4^-$ | -180.4 |
| Li$^+$(DMC$_{cc}$)BF$_4^-$ | -165.3 |
| Li$^+$(DMC$_{cc}$)$_2$BF$_4^-$ | -179.4 |
| Li$^+$(DMC$_{cc}$)$_3$BF$_4^-$ | -183.4 |
| Li$^+$(DMC$_{cc}$)PF$_6^-$ | -158.7 |
| Li$^+$(DMC$_{cc}$)$_2$PF$_6^-$ | -170.8 |
| Li$^+$(DMC$_{cc}$)$_3$PF$_6^-$ | -175.9 |

**Table 5**. Relative Gibbs Free Energies ΔG of the $Li^+(PC)_{0-3}(A^-)$, $Li^+(DMC_{cc})_{0-3}(A^-)$ and $Li^+(DMC_{ct})_{0-3}(A^-)$ clusters (where $A^- = ClO_4^-, BF_4^-, PF_6^-$) (kcal/mol).

| Fragments | ΔG=G − G$^{min}$ *(kcal/mol) | | |
|---|---|---|---|
|  | $A^- = ClO_4^-$ | $A^- = BF_4^-$ | $A^- = PF_6^-$ |
| $Li^+(A^-)$ + 3 PC | 14.9 | 17.1 | 21.2 |
| $Li^+(PC)A^-$ + 2 PC | 2.7 | 3.6 | 5.6 |
| $Li^+(PC)_2A^-$ + PC | 0.0 | 0.0 | 0.0 |
| $Li^+(PC)_3A^-$ | 2.3 | 0.8 | 6.0 |
| $Li^+(A^-)$ + 3 $DMC_{cc}$ | 13.3 | 17.2 | 12.9 |
| $Li^+(DMC_{cc})A^-$ + 2 $DMC_{cc}$ | 2.6 | 3.9 | 3.2 |
| $Li^+(DMC_{cc})_2A^-$ + $DMC_{cc}$ | 0.0 | 0.0 | 0.0 |
| $Li^+(DMC_{cc})_3A^-$ | 5.0 | 5.8 | 7.7 |
| $Li^+(A^-)$ + 3 $DMC_{ct}$ | 16.7 | 20.1 | 19.4 |
| $Li^+(DMC_{ct})A^-$ + 2 $DMC_{ct}$ | 3.6 | 5.5 | 3.9 |
| $Li^+(DMC_{ct})_2A^-$ + $DMC_{ct}$ | 0.0 | 0.0 | 0.0 |
| $Li^+(DMC_{ct})_3A^-$ | 3.8 | 3.5 | 3.8 |

* G$^{min}$ corresponds to the minimum free energy value for each of the separate data set presented in the different columns.

**Table 6**. Computed ΔG reaction values (kcal/mol) for the substitution of two solvent molecules by an Anion ($ClO_4^-$, $BF_4^-$ and $PF_6^-$) towards the $Li^+(S)_2(A^-)$ solvation structure (S= EC, PC, $DMC_{cc}$ and $DMC_{ct}$).

| Solvation Mechanism | (S)$_2$ | | | |
| --- | --- | --- | --- | --- |
| | S = EC | S = PC | S = $DMC_{cc}$ | S = $DMC_{ct}$ |
| $Li^+(S)_4 + BF_4^- \rightarrow Li^+(S)_2BF_4^- + 2S$ | -79.5 | -80.1 | -93.5 | -70.6 |
| $Li^+(S)_4 + ClO_4^- \rightarrow Li^+(S)_2ClO_4^- + 2S$ | -74.5 | -78.9 | -90.3 | -68.1 |
| $Li^+(S)_4 + PF_6^- \rightarrow Li^+(S)_2PF_6^- + 2S$ | -69.8 | -72.2 | -83.2 | -63.9 |

**TOC GRAPHIC**

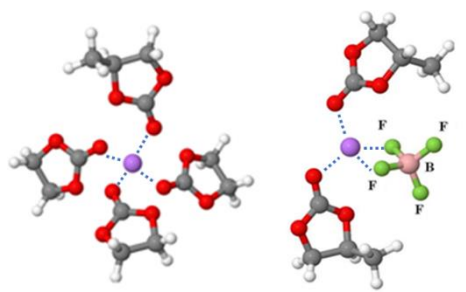

# Supplementary Information for:

# "Solvent and Salt Effect on Lithium Ion Solvation and Contact Ion Pair Formation in Organic Carbonates: A Quantum Chemical Perspective"

*Veerapandian Ponnuchamy[†,‡,||,*], Stefano Mossa[†,*] and Ioannis Skarmoutsos[†,*]*

[†] Univ. Grenoble Alpes, CEA, CNRS, INAC-SYMMES, 38000 Grenoble, France

[‡] Univ. Grenoble Alpes, CEA, LITEN-DEHT, 38000 Grenoble, France

[||] Current address: InnoRenew CoE, Izola, Slovenia
* Emails: veera.pandi33@gmail.com / Phone: +33 7 78 10 42 94, stefano.mossa@cea.fr / Phone: +33 4 38 78 35 77 / Fax: +33 4 38 78 56 91, iskarmoutsos@hotmail.com / Phone: +33 4 38 78 35 77

**Table S1.** Calculated Enthalpy and Entropy of $Li^+(EC)_n(PC)_m$ and $Li^+(PC)_n(DMC)_m$ (n + m = 4) Clusters (H in kcal/mol and S in cal mol$^{-1}$ K$^{-1}$)

| Clusters | H | S |
| --- | --- | --- |
| Mixtures EC/PC | | |
| $Li^+(EC)_3(PC)$ | -5842.1 | 167.2 |
| $Li^+(EC)_2(PC)_2$ | -6209.6 | 173.9 |
| $Li^+(EC)(PC)_3$ | -6576.9 | 180.3 |
| Mixtures PC/DMC | | |
| $Li^+(PC)_3(DMC_{cc})$ | -6734.5 | 193.1 |
| $Li^+(PC)_2(DMC_{cc})_2$ | -6524.7 | 209.5 |
| $Li^+(PC)(DMC_{cc\ 3}$ | -6314.5 | 204.7 |
| Pure Solvents | | |
| EC | -1369.3 | 72.8 |
| PC | -1736.0 | 79.6 |
| $DMC_{cc}$ | -1530.2 | 82.5 |

**Table S2.** Calculated and experimental (Ref. 27) C=O stretching frequencies (cm$^{-1}$) for PC and DMC

|  | **PC** | **DMC** | **Experiment (Ref. 27)** |
|---|---|---|---|
| Li$^+$(PC)$_4$ | 1776, 1769 |  | uncoordinated PC (1805 and 1790), **coordinated PC (1772 and 1752)** |
| Li$^+$(PC)$_3$(DMC$_{cc}$) | 1773, 1763 | 1708 |  |
| Li$^+$(PC)$_2$(DMC$_{cc}$)$_2$ | 1774, 1762 | 1697 |  |
| Li$^+$PC(DMC$_{cc}$)$_3$ | 1773, 1762 | 1686 | uncoordinated DMC (1755), and **coordinated DMC 1724 cm$^{-1}$** ) |
| Li$^+$(DMC$_{cc}$)$_4$ |  | 1678 |  |

**Table S3.** Thermodynamical Parameters of $Li^+(PC)_{1-3}(Anion-)$ and $Li^+(DMC)_{1-3}(Anion-)$ Clusters (H in kcal/mol and S in cal mol$^{-1}$ K$^{-1}$)

| Fragments | H | S |
|---|---|---|
| $Li^+(PC)ClO_4^-$ | -2299.1 | 128.2 |
| $Li^+(PC)_2ClO_4^-$ | -4048.5 | 172.5 |
| $Li^+(PC)_3ClO_4^-$ | -5793.8 | 212.9 |
| $Li^+(PC)BF_4^-$ | -2475.1 | 121.5 |
| $Li^+(PC)_2BF_4^-$ | -4225.2 | 166.1 |
| $Li^+(PC)_3BF_4^-$ | -5971.1 | 210.2 |
| $Li^+(PC)PF_6^-$ | -2611.4 | 126.1 |
| $Li^+(PC)_2PF_6^-$ | -4364.3 | 167.9 |
| $Li^+(PC)_3PF_6^-$ | -6108.3 | 200.7 |
| $Li^+(DMC_{cc})ClO_4^-$ | -2093.9 | 123.9 |
| $Li^+(DMC_{cc})_2ClO_4^-$ | -3637.2 | 170.9 |
| $Li^+(DMC_{cc})_3ClO_4^-$ | -5171.4 | 223.1 |
| $Li^+(DMC_{cc})BF_4^-$ | -2269.2 | 124.2 |
| $Li^+(DMC_{cc})_2BF_4^-$ | -3813.5 | 172.3 |
| $Li^+(DMC_{cc})_3BF_4^-$ | -5347.6 | 222.1 |
| $Li^+(DMC_{cc})PF_6^-$ | -2407.6 | 122.3 |
| $Li^+(DMC)_2PF_6^-$ | -3949.9 | 175.0 |
| $Li^+(DMC_{cc})_3PF_6^-$ | -5485.2 | 214.3 |

**Table S4.** Thermodynamical Parameters and Binding Energies of Li$^+$(PC)$_{1-3}$(A$^-$) and Li$^+$(DMC)$_{1-3}$(A$^-$) Clusters (H and BE are in kcal/mol and S in cal mol$^{-1}$ K$^{-1}$)$^a$

| Fragments | H | BE | S |
|---|---|---|---|
| DMCct | -1527.2 | | 81.3 |
| Li$^+$(DMCct)ClO$_4^-$ | -2091.6 | -163.7 | 128.7 |
| Li$^+$(DMCct)$_2$ClO$_4^-$ | -3633.8 | -178.7 | 171.5 |
| Li$^+$(DMCct)$_3$ClO$_4^-$ | -5167.4 | -185.1 | 218.7 |
| Li$^+$(DMCct)BF$_4^-$ | -2264.4 | -163.3 | 133.2 |
| Li$^+$(DMCct)$_2$BF$_4^-$ | -3809.6 | -181.2 | 173.0 |
| Li$^+$(DMCct)$_3$BF$_4^-$ | -5343.9 | -188.3 | 218.7 |
| Li$^+$(DMCct)PF$_6^-$ | -2404.6 | -158.5 | 140.8 |
| Li$^+$(DMCct)$_2$PF$_6^-$ | -3946.6 | -173.2 | 185.7 |
| Li$^+$(DMCct)$_3$PF$_6^-$ | -5480.7 | -180.1 | 231.4 |

$^a$H = Enthalpy, BE = Binding Energy, S = Entropy.

**Table S5.** Calculated free energy changes for the transitions between different solvation structures containing the cis-cis and cis-trans conformers of DMC.

| Reaction | $\Delta$G (kcal/mol) | | |
| --- | --- | --- | --- |
| | $A^-=ClO_4^-$ | $A^-=BF_4^-$ | $A^-=PF_6^-$ |
| $Li^+(DMC_{cc})_2(A^-) + DMC_{ct} \rightarrow Li^+(DMC_{cc})(DMC_{ct})(A)^- + DMC_{cc}$ | -2.2 | -2.3 | -4.1 |
| $Li^+(DMC_{cc})(DMC_{ct})(A)^- + DMC_{ct} \rightarrow Li^+(DMC_{ct})_2(A^-) + DMC_{cc}$ | -1.1 | -0.6 | -2.2 |

**FIGURES**

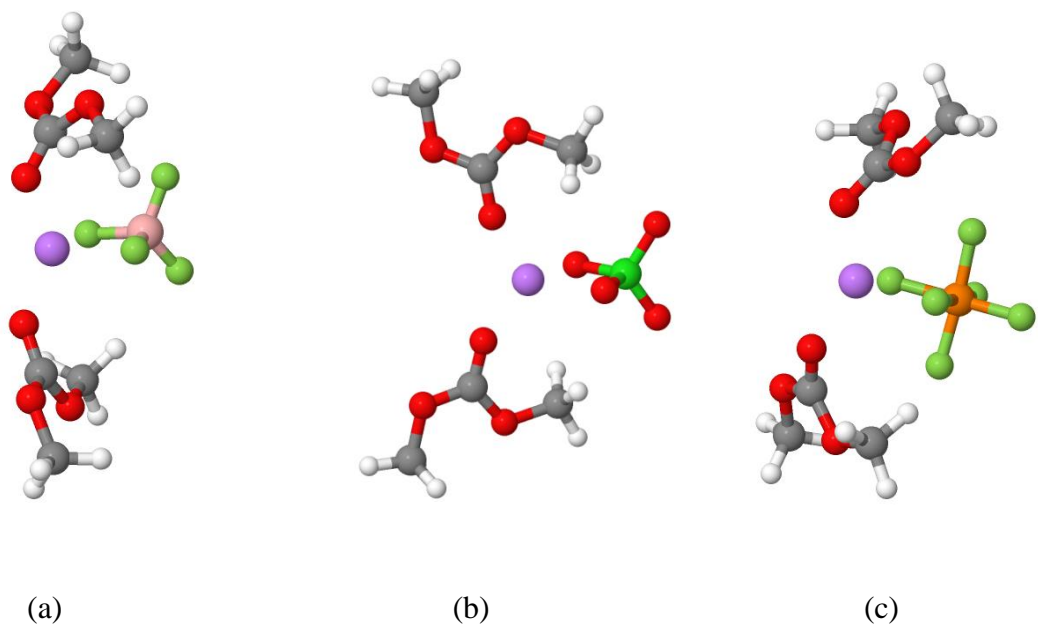

(a)            (b)            (c)

**Figure S1**: Most stable optimized geometries of (a) $Li^+(DMC_{ct})_2(ClO_4^-)$, (b) $Li^+(DMC_{ct})_2(BF_4^-)$, (c) $Li^+(DMC_{ct})_2(PF_6^-)$